\documentstyle[epsfig]{aipproc}

\begin{document}
\title{Black hole neutron star coalescence as a source of gamma--ray bursts}

\author{William H. Lee and W\l odzimierz Klu\'{z}niak$^{\dagger}$}
\address{Physics Department, University of Wisconsin \\ Madison, WI,
53706 \\ $^{\dagger}$Also Copernicus Astronomical Center, ul. Bartycka
18, 00-716 Warszawa, Poland}

\maketitle

\begin{abstract}

We present the results of hydrodynamic (SPH) simulations showing the
coalescence of a black hole with a neutron star to be a promising
theoretical source of short duration gamma-ray bursts. The favorable
features of the process include rapid onset, millisecond variability,
a duration much longer than the dynamical timescale, and a range of
outcomes sufficient to allow variety in the properties of individual
gamma-ray bursts.  Interestingly, the process of coalescence differs
rather markedly from past predictions.

\end{abstract}

\section*{Introduction}

The coalescence of a tight binary composed of two extremely compact
objects (two neutron stars or one neutron star and a black hole) has
been suggested as a possible source of cosmological GRBs \cite{bp}. In
the black hole--neutron star scenario, the neutron star was expected
to be tidally disrupted and to form a long--lived accretion torus
around the black hole, thus powering a relativistic fireball that
produces the observed GRB. Our aim is to investigate the outcome of
the coalescence from the standpoint of the hydrodynamics, and explore
how the initial mass ratio and the degree of tidal locking in the
binary affect the final outcome. In particular, it is of great
interest to determine if the neutron star is completely disrupted in a
single encounter and if a baryon--free axis persists throughout the
process.

\section*{Numerical method and assumptions}

For the calculations presented here we have used a three dimensional
newtonian smooth particle hydrodynamics \cite{monaghan} code
\cite{acta}. The neutron star is modeled via a stiff polytropic
equation of state, i.e $P=K\rho^{\Gamma}$, with $\Gamma=3$.  The
unperturbed radius of the spherical neutron star is $R_{NS}=13.4$km
and its mass is $M_{NS}=1.4$ M$_{\odot}$. The black hole is modeled as
a Newtonian point mass with an absorbing boundary at the Schwarschild
radius $r=2GM_{BH}/c^{2}$. Any particle in the simulation that crosses
this boundary is absorbed by the black hole and its mass is added to
that of the black hole.  For the different simulations we vary the
initial mass ratio of the binary $q=M_{NS}/M_{BH}$ by adjusting the
mass of the black hole only.

\section*{Results}

\subsection*{Tidally locked binaries}

For any value of the mass ratio one can construct tidally locked
initial configurations that are in equilibrium \cite{rasio},
provided the binary separation is large enough so that no mass
transfer occurs. If the neutron star fills its Roche lobe, any further
decrease in separation will produce mass transfer, which can be stable
or unstable, depending on the initial mass ratio. We study the
coalescence of the binary by performing dynamical runs starting with
initial configurations that are on the verge of initiating mass
transfer.

\subsubsection*{High mass ratios}

\begin{figure}[h!] 
\centerline{\epsfig{file=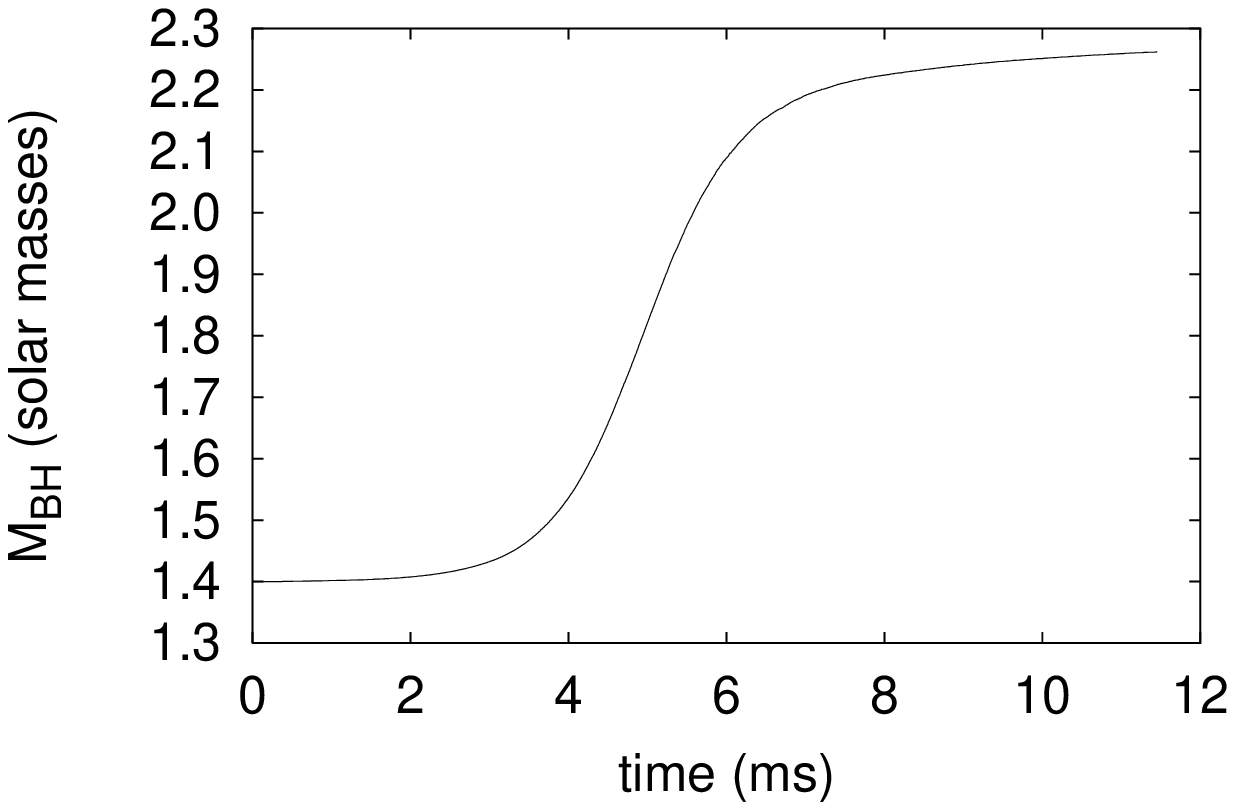,height=1.66in,width=2.5in}\epsfig{file=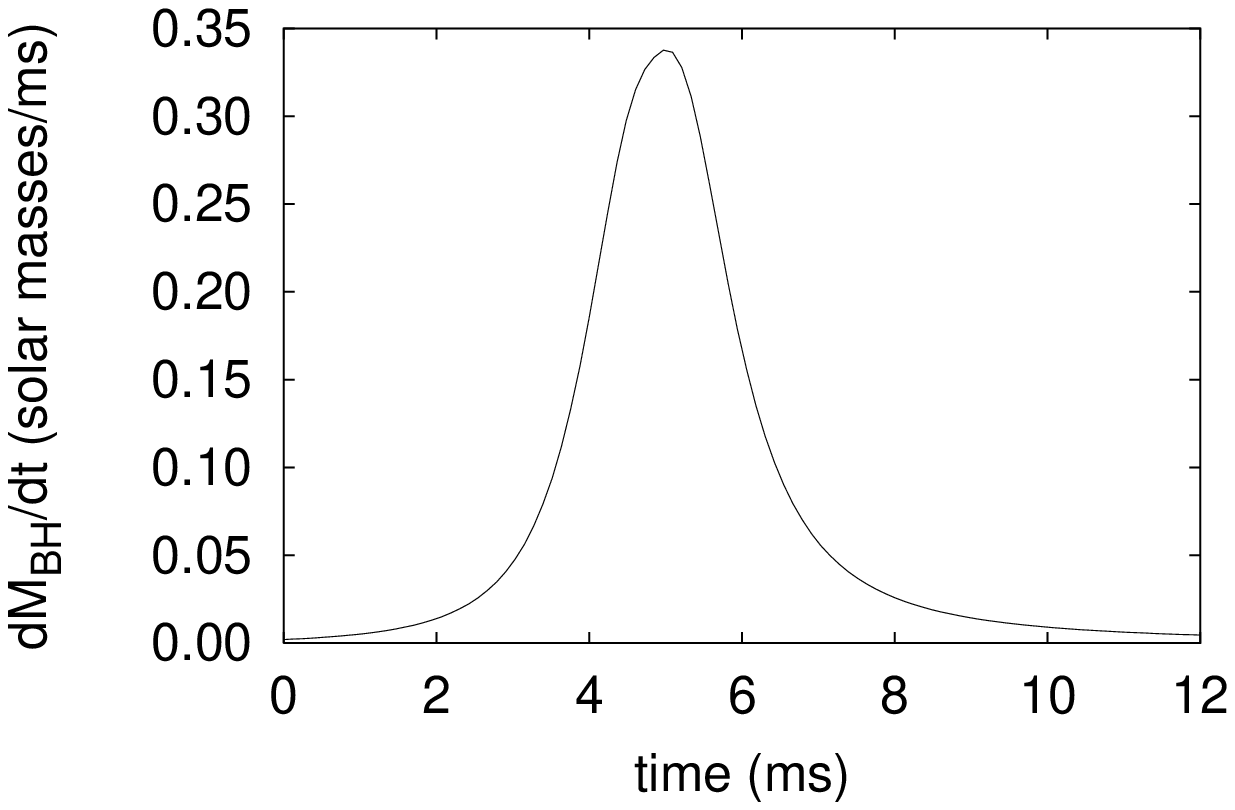,height=1.66in,width=2.5in}}
\vspace{10pt}
\caption{Black hole mass (left) and accretion rate onto the black hole
(right) for an initial mass ratio of $q=1$ in a tidally locked
binary.}
\label{F:M-O4:1}
\end{figure}

\begin{figure}[h!] 
\centerline{\epsfig{file=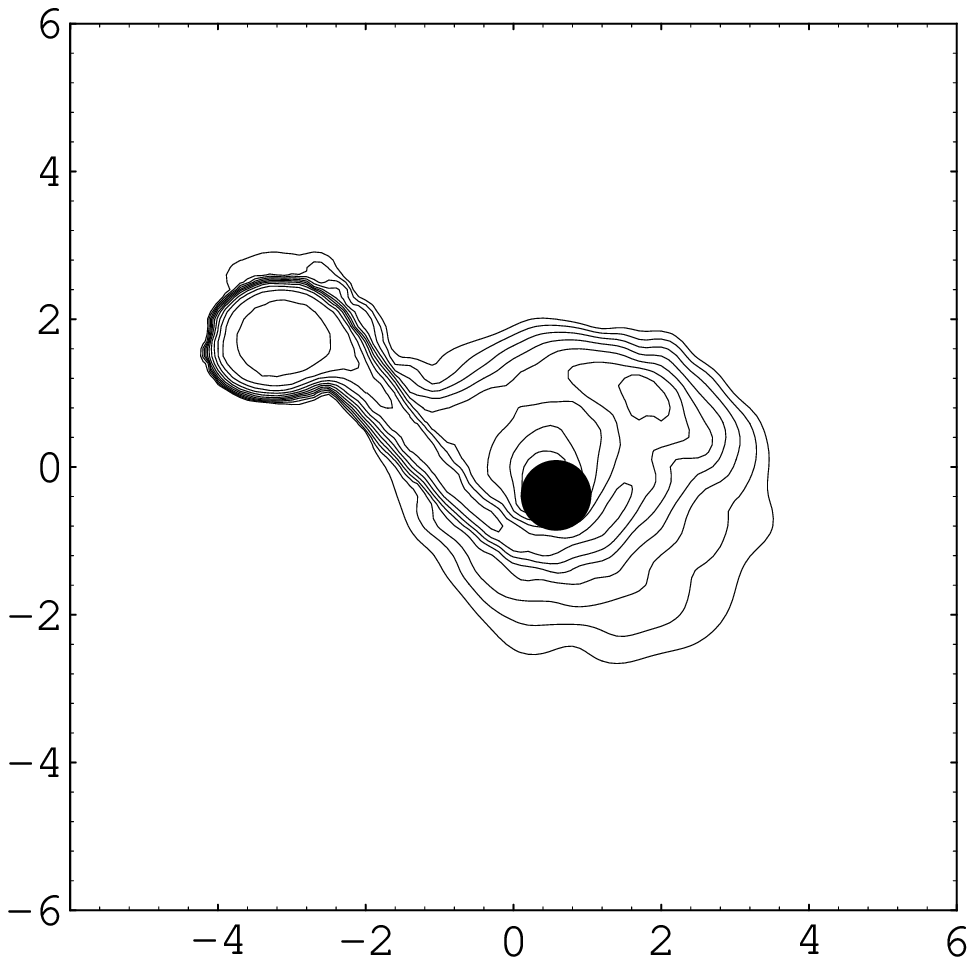,height=2.5in,width=2.5in}\epsfig{file=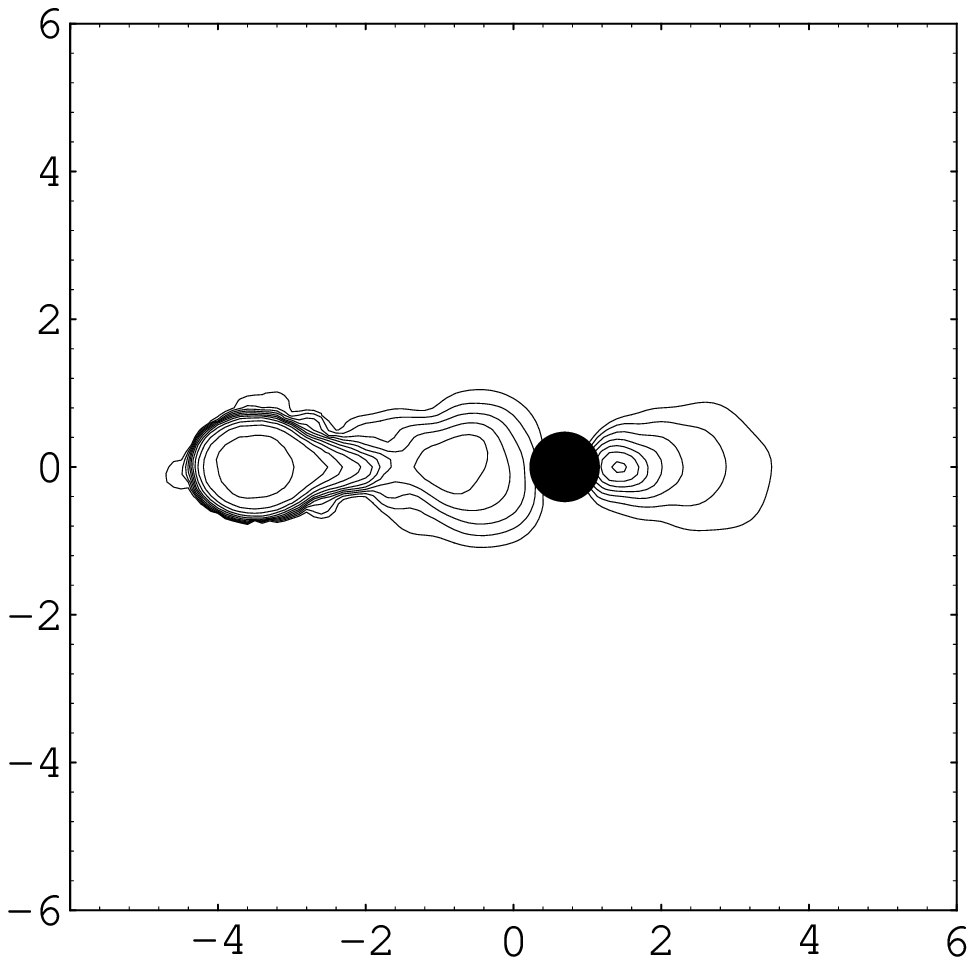,height=2.5in,width=2.5in}}
\vspace{10pt}
\caption{Density contours for the initially tidally locked binary with
$q=1$ in the orbital plane (left) and a meridional plane
(right). Orbital rotation is counterclockwise. There are eleven evenly
spaced logarithmic contours between $ 5 \times 10^{14}$~kg~m$^{-3}$
and $5 \times 10^{17}$~kg~m$^{-3}$. The axes are labeled in units of $R_{NS}$.}
\label{F:M-O4:2}
\end{figure}

For an initial mass ratio of unity, the neutron star overflows its
Roche lobe at a separation $r=2.78R_{NS}$ (the orbital period at this
separation is 2.3 ms). A very fast episode of mass transfer ensues, in
which 0.9 M$_{\odot}$ are accreted by the black hole (Figure \ref{F:M-O4:1}).  
Note that we have not included a gravitational backreaction force in
our calculations, but the orbital decay is driven by hydrodynamical
effects which are comparable in magnitude to angular momentum losses
to gravitational radiation. We observe the formation of a transient
accretion torus (Figure \ref{F:M-O4:2}) containing about 0.1 solar
masses around the black hole, lasting for several orbits.
To the limit of our
resolution ($10^{-4}$ M$_{\odot}$), there is a baryon--free axis along
the rotation axis of the binary throughout the simulation. We are
unable at present to follow the further evolution of the matter spread
around the black hole for more than 11 ms. Note that the neutron star
is not completely disrupted by the initial encounter, but that a
low--mass cores survives and is transferred to a higher orbit by
conservation of angular momentum. This core contains approximately
0.43 solar masses, and thus the final mass ratio in the binary is
$q=0.19$, while the separation has increased to about 47 km.

\subsubsection*{Lower mass ratios}

\begin{figure}[h!] 
\centerline{\epsfig{file=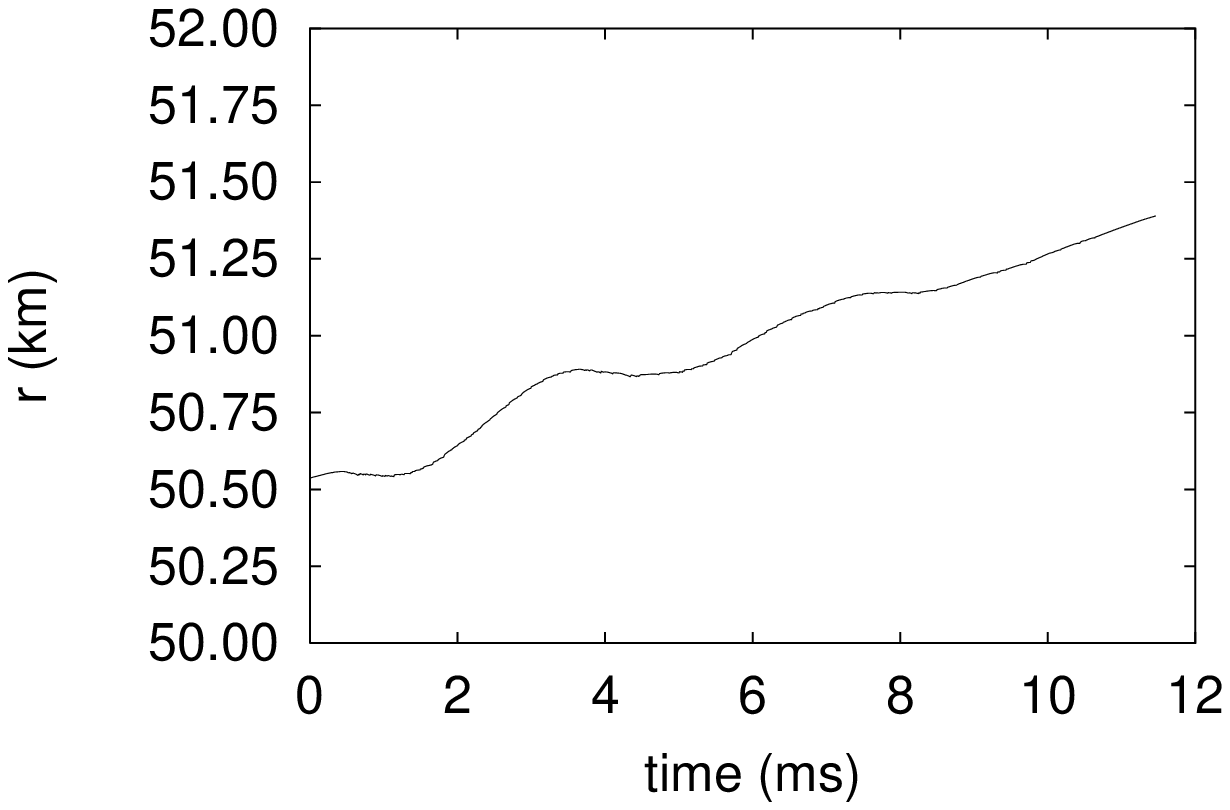,height=1.66in,width=3.0in}}
\vspace{10pt}
\caption{Binary separation $r$ as a function of time for an initially
tidally locked binary with an initial mass ratio of $q=0.31$.}
\label{F:M-O4:3}
\end{figure}

We have also investigated the outcome of a coalescence for lower mass
ratios, i.e. with a more massive black hole. For an initial mass ratio
of $q=0.31$, corresponding to a black hole mass of 4.5 M$_{\odot}$,
the evolution of the binary is quite different than that presented
above. The separation corresponding to the onset of mass transfer is
$r=50.4$ km, but for this case the neutron star is not violently
disrupted. Instead, mass transfer through Roche--lobe overflow occurs,
with about 1\% of the mass of the neutron star being transferred to
the black hole. The binary separation increases slightly (Figure
\ref{F:M-O4:3}), and the whole episode conserves total orbital angular
momentum.
All the matter stripped from the neutron star is accreted by the black
hole, hence no torus forms, and a baryon--free axis is also
present. However, the mass transfer episode lasts about 11 ms, and
angular momentum losses to gravitational radiation cannot be
ignored. Also, the assumption of tidal locking is not thought to be
realistic \cite{bild}.

\subsection*{Binaries which are not tidally locked}

\begin{figure}[h!] 
\centerline{\epsfig{file=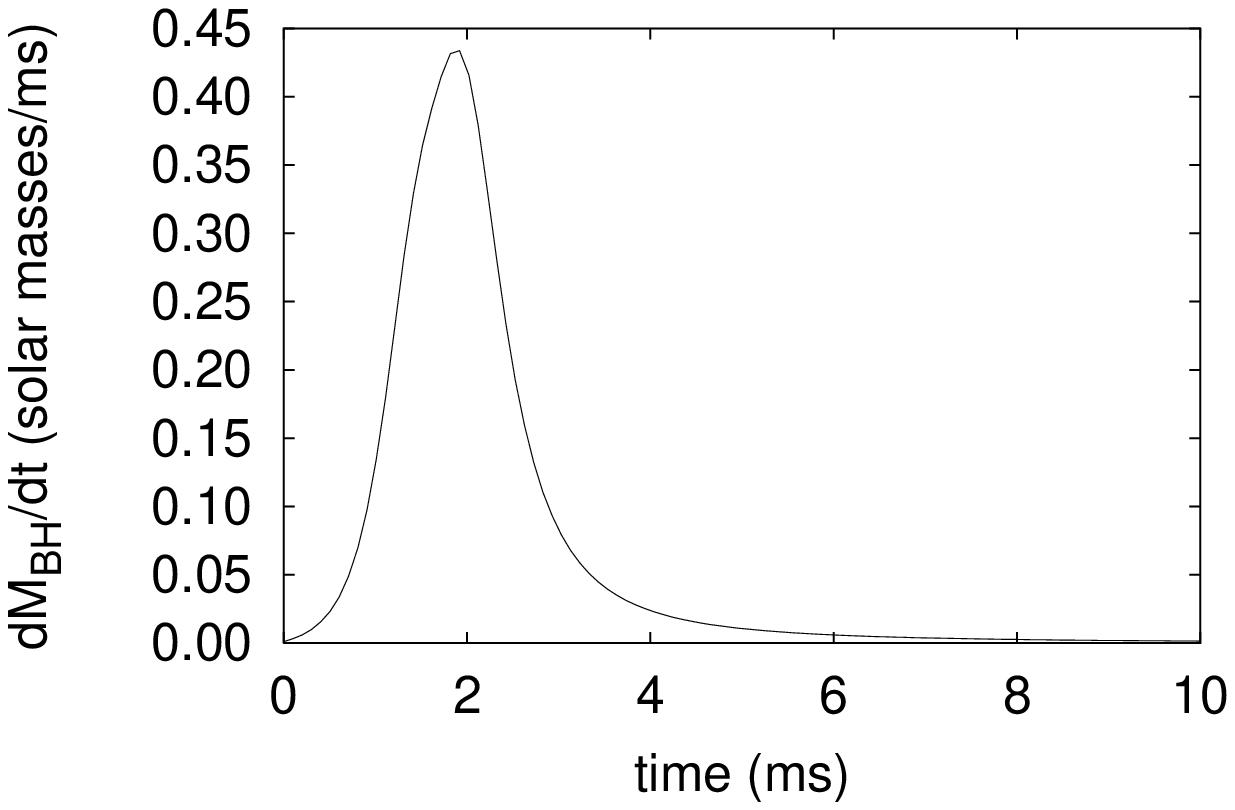,height=1.66in,width=2.5in}}
\vspace{10pt}
\caption{Mass accretion rate onto the black hole for a binary with an
initial mass ratio $q=0.31$ that is not tidally locked (the neutron
star is initally not spinning).}
\label{F:M-O4:4}
\end{figure}

Suppose we now remove the restriction of tidal locking. If the neutron
star is initially spherical and non--rotating, the presence of the
black hole will create a tidal bulge on the neutron star. The neutron
star will then be spun--up to some degree, and this spin angular
momentum will be extracted from the orbital component. The orbit will
thus decay on a shorter time scale. We have performed just such a
calculation, starting with an initial mass ratio of $q=0.31$ as above,
and with a neutron star that is initially spherical and not
spinning. The orbital evolution is now much more rapid, with an
episode of mass transfer lasting less than 4~ms (Figure \ref{F:M-O4:4}).
The total mass transferred is 0.6 M$_{\odot}$, and again the core of
the neutron star (0.8 M$_{\odot}$) is not disrupted, but transferred
to a higher orbit. The final separation is approximately 70 km.

\section*{Discussion}

Our simulations \cite{apjl,longpaper} lead us to believe that for the
equation of state used here the survival of the neutron star core is a
robust result. The timescale for the entire coalescence is thus
lenghtened from a few milliseconds to at least several tens of
milliseconds, since a binary with a lower mass ratio and greater
separation (the result of the initial episode of mass transfer) will
take a longer time to decay via angular momentum losses to
gravitational radiation. Furthermore, to the limit of our resolution,
there is a baryon--free axis along the rotation axis of the binary
present throughout the simulation in every case. Calculations with
improved resolution are required to determine if the baryon loading is
low enough to accomodate the requirements of the blast--wave model for
GRBs \cite{rees}.

We note that without the formation of a torus, it is difficult to
extend the time scale of the coalescence to many seconds, but we
nevertheless believe that the coalesence of a neutron star with a
black hole is a promising candidate source for the central engine of
the shorter gamma ray bursts in the bimodal distribution
\cite{kouveliotou}.

These results also suggest that black hole--neutron star binaries
might well be the production sites for low--mass neutron stars
unstable to explosion. The details of how the remnant core would react
to the violent episode of mass loss depend on the equation of state,
but it has been shown \cite{colpi,sumiyoshi} that if the mass were to
drop below the minimum required for stability, a violent explosion
would ensue. The timescale for this event is not certain, but
estimates range from a few milliseconds to several tens of seconds.

This work was supported in part by Poland's Committee for Scientific
Research under grant KBN 2P03D01311 and by DGAPA--UNAM.


\begin{references}

\bibitem{bild} Bildsten, L., Cutler, C., {\it Astrophys. J.}\ {\bf
400}, 175 (1992)

\bibitem{colpi} Colpi, M., Shapiro, S.L., Teukolsky, S.A., {\it
Astrophys. J.}\ {\bf 369}, 422 (1991)

\bibitem{apjl} Klu\'zniak, W., Lee, W.H., {\it Astrophys. J. Lett.}\
submitted (1997)

\bibitem{kouveliotou} Kouveliotou, C. {\it et al.}, 1995, in {\it
Gamma Ray Bursts}, C. Kouveliotou, M.F. Briggs, G.J. Fishman,
eds. (AIP, New York), 42 (1995)

\bibitem{acta} Lee, W.H., Klu\'zniak, W. 1995, {\it Acta Astron.}\
 {\bf45}, 705 (1995)

\bibitem{longpaper} Lee, W.H., Klu\'zniak, W., in preparation (1997)

\bibitem{rees} M\'esz\'aros, P., Rees, M.J.,
 {\it Astrophys. J.}\ {\bf 405}, 278 (1993)

\bibitem{monaghan} Monaghan, J.J., {\it Ann. Rev. Ast. \& Astrophys.}\
 {\bf 30}, 543 (1992).

\bibitem{bp} Paczy\'nski, B., {\it Acta Astron.}\ {\bf 41}, 257 (1991)

\bibitem{rasio} Rasio, F., Shapiro, S.L., {\it Astrophys. J.}\ {\bf
432}, 242 (1994)

\bibitem{sumiyoshi} Sumiyoshi, K., Yamada, S.,
Suzuki, H., Hilldebrandt, W., 1997, preprint, astro-ph/9707230.

\end{references}
\end{document}